\documentclass[aps,prd,twocolumn,superscriptaddress,longbibliography,nofootinbib]{revtex4-2}
\usepackage[utf8]{inputenc}
\usepackage{color}
\usepackage{graphicx}
\usepackage{subfigure}

\usepackage{amsmath,amssymb,amsfonts}

\def\prl{Phys. Rev. Lett.}
\def\prd{Phys. Rev. D}
\def\cqg{Class. Quantum Grav.}

\def\pau_p{Prog. Theor. Phys.}

\def\physrep{Phys. Rep.}

\def\At{{\mathcal A_{\rm T}}}
\def\Ab{{\mathcal A_{\rm B}}}

\def\erf{{\rm erf}}

\begin{document}

\title{Comparison of linear Brill and Teukolsky waves}

\author{Isabel Su\'arez Fern\'andez}
\affiliation{
  Centro de Astrof\'{\i}sica e Gravita\c c\~ao -- CENTRA,
  Departamento de F\'{\i}sica, Instituto Superior T\'ecnico -- IST,
  Universidade de Lisboa -- UL, Av.\ Rovisco Pais 1, 1049-001 Lisboa,
  Portugal}

\author{Thomas W. Baumgarte}
\affiliation{Department of Physics and Astronomy, Bowdoin College, Brunswick, ME 04011, USA}

\author{David Hilditch}
\affiliation{
  Centro de Astrof\'{\i}sica e Gravita\c c\~ao -- CENTRA,
  Departamento de F\'{\i}sica, Instituto Superior T\'ecnico -- IST,
  Universidade de Lisboa -- UL, Av.\ Rovisco Pais 1, 1049-001 Lisboa,
  Portugal}

\begin{abstract}
Motivated by studies of critical phenomena in the gravitational collapse of vacuum gravitational waves we compare, at the linear level, two common approaches to constructing gravitational-wave initial data.  Specifically, we construct analytical, linear Brill wave initial data and compare these with Teukolsky waves in an attempt to understand the different numerical behavior observed in dynamical (nonlinear) evolutions of these two different sets of data.  In general, the Brill waves indeed feature higher multipole moments than the quadrupolar Teukolsky waves, which might have provided an explanation for the differences observed in the dynamical evolution of the two types of waves.  However, we also find that, for a common choice of the Brill-wave seed function, all higher-order moments vanish identically, rendering the (linear) Brill initial data surprisingly similar to the Teukolsky data for a similarly common choice of its seed function.    
\end{abstract}

\maketitle

\section{Introduction}
\label{sec:intro}

Critical phenomena in gravitational collapse, first reported by Choptuik \cite{Cho93}, refer to properties of solutions to Einstein's equations close to the threshold of black-hole formation.  Specifically, Choptuik considered spherically symmetric massless scalar fields minimally coupled to Einstein's equations.  Evolving several one-parameter families of initial data numerically, he observed the existence of a {\em critical parameter} $\eta_*$ that separates {\em supercritical} data, which ultimately form a black hole, from {\em subcritical} data, which do not.  Critical phenomena, with surprising resemblance to similar phenomena in other fields of physics, emerge close to the critical parameter $\eta_*$.  For super-critical data, for example, the black-hole mass will follow an approximate power-law
\begin{equation} \label{M_scaling}
    M \simeq (\eta - \eta_*)^\gamma,
\end{equation}
where the critical exponent $\gamma$ depends on the matter-model, but not on the initial data.  For massless scalar fields, for example, Choptuik found $\gamma \simeq 0.37$.  Also, close to criticality the evolution of the initial data will, at intermediate times, follow a {\em self-similar critical} solution with, depending on the matter model, either discrete or continuous self-similarity. 

Following Choptuik's original announcement, numerous authors have studied critical phenomena in gravitational collapse, both numerically and analytically, for a number of different matter models, symmetry assumptions, and asymptotics (see \cite{Gun03,GunM07} for reviews).  As a result of these studies, critical collapse is now well understood in the context of spherical symmetry.  For example, the power-law scaling for dimensional quantities, like the mass in (\ref{M_scaling}), can be explained from perturbations of a unique self-similar critical solution, with the inverse of the Lyapunov exponent of a single unstable mode yielding the critical exponent $\gamma$  (see, e.g., \cite{KoiHA95,Mai96,Gun97}).  

The situation is much less clear in the absence of spherical symmetry, which includes what is perhaps the most intriguing case of critical collapse, namely the gravitational collapse of vacuum gravitational waves.  Critical phenomena in this collapse were first reported by \cite{AbeE93,AbeE94}, but, despite significant effort by a number of authors (see, e.g., Table I in \cite{Hiletal13} for a summary of attempts), it has been difficult to reproduce these results (but see \cite{HilWB17,LedK21} for recent progress).  Some of the problems associated with these calculations appear to be numerical in nature, but others may also be conceptual issues that arise in the absence of spherical symmetry (see also \cite{SuaVH21}).

Different authors have adopted different types of initial data for simulations of vacuum gravitational waves.  One type of initial data are often called ``Teukolsky waves" (\cite{Teu82}, see Section \ref{sec:indata:teukolsky} below).  These data represent quadrupolar, linear perturbations of the Minkowski spacetime, which can be ``dressed up" in different ways to yield non-linear solutions to Einstein's constraint equations.  A second type of initial data are so-called ``Brill waves" (\cite{Bri59}, see Section \ref{sec:indata:brill} below).  Constructing Brill waves entails solving one elliptic equation (see eq.~\ref{Ham1} below), whose solution then provides a nonlinear vacuum solution to Einstein's constraint equations.  Both Teukolsky and Brill waves allow for an arbitrary ``seed function", for which many authors have adopted Gaussian profiles.

One of the mysteries emerging from the study of critical collapse of gravitational waves is that the above types of initial data appear to behave differently when evolved numerically.  The authors of \cite{LedK21} report that different initial data will result in different critical exponents $\gamma$, and hence, presumably, different threshold solutions.  Several authors have also observed that the evolution of Brill wave initial data is less stable numerically than that for Teukolsky waves; the authors of \cite{Hiletal13} pose the question ``Why is it so difficult to evolve Brill wave data?", while the authors of \cite{LedK21} report that ``those [data] most defying our bisection attempts were the Brill initial data".  It would therefore be desirable to gain some understanding of what characteristics distinguish the two types of initial data, and how they affect the dynamical evolution.

In an independent approach to exploring the effects of the absence of spherical symmetry, the authors of \cite{BauGH19,PerB21} studied the critical gravitational collapse of electromagnetic waves.  In this case the initial data are constructed by adopting an axisymmetric spherical electromagnetic wave (which is linear by nature) of a given multipole moment $\ell$, at the moment of time-symmetry, and then solving Einstein's constraint equations.  The non-linear terms in Einstein's equations will couple different multipoles, of course, but one nevertheless expects the initial data to be dominated by the linear ``seed" data.  According to these studies, initial data for different multipole moments will result in qualitatively different threshold solutions, suggesting the absence of a unique critical solution.  For dipole data, with $\ell = 1$, for example, the authors of \cite{BauGH19} found a center of collapse at the origin, while for quadrupole data, with $\ell = 2$, the authors of \cite{PerB21} found two separate centers of collapse on the axis of symmetry.  The latter is consistent with the findings of \cite{HilWB17,LedK21} for gravitational waves. Moreover, the evolution of higher-order multipoles appears to be increasingly difficult, even apart from the need for higher angular resolution.

This latter observation suggests a possible explanation for the differences in the evolution between the Brill and Teukolsky gravitational-wave initial data, namely in terms of multiple moments.  In this paper we therefore construct analytical, linear solutions describing small-amplitude Brill waves, and compare these directly with Teukolsky waves.  We compare the resulting data, for given choices of the seed functions, in three different ways: (i) we compare the data directly by transforming the Brill data to the TT-gauge of the Teukolsky data (Section \ref{sec:transform}), (ii) we compute the gauge-invariant Moncrief functions of different orders $\ell$ (see Section \ref{sec:moncrief}), and (iii) we compute and compare the (gauge-invariant) Kretschmann scalar (Section \ref{sec:kretschmann}).  As we will find in Section \ref{sec:compare} below, linearized Brill data are, in general, linear combinations of different multipole moments, and may therefore be more complicated to evolve than Teukolsky data, which are purely quadrupolar by construction.  To our surprise, however, all multipoles higher than quadrupole vanish {\em exactly} for a common choice for the Brill-data seed function (see \cite{HolMWW93}).  For this choice, the two sets of initial data are in fact quite similar qualitatively (assuming a Gaussian seed function for the Teukolsky waves).  We therefore conclude that the root causes for their differences in nonlinear evolution probably cannot be found at the linear level, at least not in terms of the multipole moments.

\section{Linear gravitational-wave initial data}
\label{sec:indata}

\subsection{Teukolsky waves}
\label{sec:indata:teukolsky}

Quadrupolar vacuum gravitational-wave solutions to the linearized Einstein equations are commonly referred to as Teukolsky waves \cite{Teu82} (see also \cite{Rin08} for a generalization to higher multipoles, as well as \cite{ShaB10} for a textbook treatment).  Using geometrized units with $c = 1$  the metric, expressed in transverse-traceless gauge (see eqs.~\ref{TT} below), may be written in the form
\begin{align}
    ds^2  & = -dt^2 + dr^2 \, \big\{ 1 + A f_{rr} \big\} 
    + r \, dr d\theta \, \big\{ 2Bf_{r\theta} \big\} +  \nonumber \\
    &  r \sin(\theta) \, dr d\phi \, \big\{ 2Bf_{r\phi} \big\}  + 
    r^2 \, d\theta^2 \, \big\{ 1 + Cf^{(1)}_{\theta\theta} + Af^{(2)}_{\theta\theta} \big\}+  \nonumber \\
    &  r^2 \sin(\theta) \, d\theta d\phi \, \big\{ 2(A-2C)f_{\theta\phi} \big\} +  \nonumber \\
&    r^2\sin^2 (\theta) \, d\phi^2 \, \big\{ 1 + Cf^{(1)}_{\phi\phi} + Af^{(2)}_{\phi\phi} \big\},
    \label{eq:ds_Teuk}
\end{align}
where the $f_{ij}$ are angular functions (which, for $\ell = 2$ and $m = 0$, we list in eq.~\ref{eq:A_ang_fcts} of Appendix \ref{app:metric}) and the the coefficients $A$, $B$, and $C$ can be constructed from a seed function $F(t,r)$ (see, e.g., Section 9.1.2 in \cite{ShaB10} for details).  A common choice for this seed function is a linear combination of Gaussians
\begin{align}\label{eqn:Teuk_Seed}
F(t,r) = \At \lambda ^4 \Big( & (t-r) \,  e^{-((r-t)/\lambda)^2} - \nonumber \\ 
&  (r+t) \, e^{-((r+t)/\lambda)^2} \Big),
\end{align}
for which $t = 0$ becomes a moment of time symmetry.  In (\ref{eqn:Teuk_Seed}) the dimensionless constant $\At$ parametrizes the amplitude of the wave, while $\lambda$, a constant with units of length, determines its wavelength.   Adopting this seed function for axisymmetric data with $m = 0$, the functions $A$, $B$, and $C$ take the form given by (\ref{eq:A_coef_Teuk}) and the metric (\ref{eq:ds_Teuk}), evaluated at $t = 0$, becomes
\begin{align}
   ds^2 & =  -dt^2 + dr^2 \Big\{ 1 + \At \left(72 \sin ^2(\theta) -48\right) e^{-(r/\lambda)^2}\Big\}+  \nonumber \\ 
   &  r^2 d\theta^2 \Big\{ 1 +  \nonumber \\ 
   & ~~~~~  24 \At  \left(\sin ^2(\theta)  \left(-\frac{r^4}{\lambda ^4}+\frac{4 r^2}{\lambda
   ^2}-3\right) + 1 \right) e^{-(r/\lambda)^2} \Big\}  +
   \nonumber \\
   &  r d\theta dr \Big\{ 48 \At
   \sin (\theta) \cos (\theta)   \left(3 - 2 \frac{r^2}{\lambda^2} \right) e^{-(r /\lambda)^2} \Big\}+ \nonumber \\ 
   &  r^2 \sin^2 (\theta) d\phi ^2 \Big\{ 1 + \nonumber \\ 
   & ~~~~~ 24 \At  
   \left( \sin ^2(\theta)  \left( \frac{r^4}{\lambda^4} - \frac{4 r^2}{\lambda^2} \right) +1 \right)e^{-(r /\lambda)^2} \Big\}.
   \label{eq:metric_teukolsky}
\end{align}

Note that, as a vacuum solution at a moment of time symmetry, the metric (\ref{eq:metric_teukolsky}) satisfies even the nonlinear momentum constraint of Einstein's constraint equations identically, so that constructing valid (nonlinear) gravitational-wave initial data from (\ref{eq:metric_teukolsky}) requires solving the Hamiltonian constraint only.  This can be accomplished, for example, by adopting the spatial part of the metric (\ref{eq:metric_teukolsky}) as a conformally related metric in the Hamiltonian constraint, solving this equation for the conformal factor, and then constructing a new spatial metric from the two.  

In this paper, however, we focus on linear data only.  In particular, we read off the spatial ``Teukolsky" metric $\gamma_{ij}^{\rm T}$ from (\ref{eq:metric_teukolsky}), and identify from these the metric perturbations
\begin{equation} \label{eq:h_teukolsky}
    h^{\rm T}_{ij} = \gamma_{ij}^{\rm T} - \eta_{ij},
\end{equation}
where $\eta_{ij}$ is the flat metric (here in spherical polar coordinates).  The coefficients $h^{\rm T}_{ij}$ can be read off from the metric (\ref{eq:metric_teukolsky}) above, but we also list them in Appendix \ref{app:metric} for completeness.  

\subsection{Brill waves}
\label{sec:indata:brill}

Alternatively, fully non-linear, axisymmetric vacuum gravitational-wave initial data can also be constructed following the procedure suggested by Brill \cite{Bri59}.   Specifically, the spatial line element for such ``Brill waves" at a moment of time symmetry is assumed to take the form
\begin{equation} \label{metric}
    \gamma_{ij} dx^i dx^j = \psi^4 \left( e^{2 q} ( dr^2 + r^2 d \theta^2)
    + r^2 \sin^2 (\theta) d \varphi^2 \right)
\end{equation}
where $q = q(r,\theta)$ is a seed function.  Following Holz {\it et al.}~\cite{HolMWW93} as well as numerous other authors we will adopt the choice
\begin{align} \label{seed1}
    q(r,\theta) & = \Ab r^2 \sin^2 (\theta) \sigma^{-2} e^{- (r / \sigma)^2 } \nonumber \\
    & = \Ab \rho^2 \sigma^{-2} e^{-(\rho^2 + z^2) / \sigma^2}, 
\end{align}
where $\Ab$ is again an amplitude and $\sigma$ a measure of the wavelength.  Expressing the angular dependence of $q(r,\theta)$ in terms of spherical harmonics we may also write the seed function (\ref{seed1}) as 
\begin{equation} \label{q_decomp}
q(r,\theta) = q_{00}(r) Y_{00}(\theta) + q_{20}(r) Y_{20}(\theta),
\end{equation}
where
\begin{subequations} \label{q_decomp_coef}
\begin{align}
q_{00}(r) & = \sqrt{\pi} \, \frac{4 \Ab}{3} \left( \frac{r}{\sigma} \right)^2 e^{-(r/\sigma)^2}  \\
q_{20}(r) & = - \sqrt{\frac{\pi}{5}} \, \frac{4 \Ab}{3} \left( \frac{r}{\sigma} \right)^2 e^{-(r/\sigma)^2}. 
\end{align}
\end{subequations}
Not surprisingly, the expansion of the axisymmetric function $q$ requires the spherical harmonics with $m = 0$ only, which, in turn, depend on $\theta$ only.

Given the assumption of time-symmetry, the momentum constraint is satisfied identically, and the Hamiltonian constraint can be shown to take the form 
\begin{equation} \label{Ham1}
    \nabla^2 \psi = - \frac{\psi}{4} \tau,
\end{equation}
where the function $\tau = \tau(r,\theta)$ is given by
\begin{equation}
\tau \equiv \frac{\partial^2 q}{\partial \rho^2} + \frac{\partial^2 q}{\partial z^2}
\end{equation}
and where $\nabla^2$ denotes the flat Laplace operator.  For our choice (\ref{seed1}) we have
\begin{align} \label{tau}
    \tau(r,\theta) =  \frac{2 \Ab}{\sigma^6} e^{-(r/\sigma)^2} \Big( & 2 r^4 - 6 r^2 \sigma^2 + \sigma^4 -\nonumber \\
    &  2 r^2 (r^2 - 3 \sigma^2) \cos^2(\theta) \Big),
\end{align}
which we may express as
\begin{equation} \label{tau_decomp}
\tau(r,\theta) = \tau_{00}(r) Y_{00}(\theta) + \tau_{20}(r) Y_{20}(\theta)
\end{equation}
with
\begin{subequations} \label{tau_decomp_coef}
\begin{align}
\tau_{00}(r) & = \sqrt{\pi } \frac{4  \Ab }{3 \sigma ^6} e^{-(r / \sigma)^2 }  \left( 4 r^4  -12 r^2 \sigma ^2+ 3 \sigma ^4\right) \\
\tau_{20}(r) & = -\sqrt{ \frac{\pi}{5}} \frac{16  \Ab  }{3 \sigma ^6}e^{-( r / \sigma) ^2}\left(r^4-3 r^2 \sigma ^2\right) .
\end{align}
\end{subequations}

In general, the Hamiltonian constraint (\ref{Ham1}) does not permit analytical solutions, and therefore has be solved numerically.  For our purposes of a direct comparison with the (linear) Teukolsky waves of Section \ref{sec:indata:teukolsky}, however, it is sufficient to consider linear solutions to (\ref{Ham1}).  Towards that end we write the conformal factor as
\begin{equation}
    \psi = 1 + u,
\end{equation}
in which case the Hamiltonian constraint (\ref{Ham1}) becomes
\begin{equation} \label{Ham_lin}
    \nabla^2 u = - \frac{1}{4} \tau
\end{equation}
to linear order in the amplitude $\Ab$.  Similarly,  the line element (\ref{metric}) becomes 
\begin{align} \label{metric_lin}
    \gamma_{ij} dx^i dx^j = ~ & dr^2 + r^2 ( d \theta^2 + \sin^2 \theta d \varphi^2) +  \\
    & (4 u + 2 q) ( dr^2 + r^2 d \theta^2)
    + 4 u r^2 \sin^2 (\theta) d \varphi^2 \nonumber 
\end{align}
to linear order.

Using the Green function $G({\bf r},{\bf r}') = 1/|{\bf r} - {\bf r}'|$ we may write the solution to the linear Hamiltonian constraint (\ref{Ham_lin}) as
\begin{equation} \label{u}
    u(r,\theta,\varphi) = \frac{1}{16 \pi} \int \frac{\tau(r',\theta',\varphi') d^3x'}{| {\bf r} - {\bf r}'|}.
\end{equation}
We now expand the Green function as 
\begin{equation} \label{green_expand}
    \frac{1}{| {\bf r} - {\bf r}'|} = \frac{4 \pi}{r_>}
    \sum_{\ell,m} \frac{1}{2 \ell + 1} \frac{r_<^\ell}{r_>^\ell} Y^*_{\ell m}(\theta',\varphi') Y_{\ell m}(\theta, \varphi),
\end{equation}
where $r_>$ ($r_<$) is the greater (smaller) of the two radii $r$ and $r'$, and insert this together with (\ref{tau_decomp}) into (\ref{u}) to find 
\begin{align}
u(r,\theta,\varphi) = &  \,\frac{1}{4} \int \Big( \tau_{00}(r') Y_{00}(\theta',\varphi') + \tau_{20}(r') Y_{20}(\theta',\varphi') \Big) \nonumber \\
 &	\sum_{\ell,m} \frac{1}{2 \ell + 1} \frac{r_<^\ell}{r_>^{\ell+1}} Y^*_{\ell m}(\theta',\varphi') Y_{\ell m}(\theta,\varphi) d^3 x'.
\end{align}
We now write the volume element as $d^3x' = r'^2 dr' d\Omega'^2$ and carry out the angular integration using the orthogonality of the spherical harmonics,
\begin{equation}
\int Y^*_{\ell m}(\theta', \varphi') Y_{\ell' m'}(\theta', \varphi') d \Omega'^2 = \delta_{\ell,\ell'} \delta_{m,m'},
\end{equation}
to obtain
\begin{equation} \label{u_decomp}
u(r,\theta) = u_{00}(r) Y_{00} + u_{20}(r) Y_{20}
\end{equation}
with
\begin{subequations} \label{u_integrals}
\begin{align}
u_{00}(r) & = \frac{1}{4} \int_0^{\infty} \frac{\tau_{00}(r')\, r'^2 dr'}{r_>}  \\
u_{20}(r) & = \frac{1}{4} \int_0^{\infty} \frac{r^2_< \tau_{20}(r') \, r'^2 dr'}{r^3_>} .
\end{align}
\end{subequations}
The integrals in equations (\ref{u_integrals}) have to be split into two parts in order to account for $r'$ being either smaller or greater than $r$, e.g.
\begin{equation}
u_{00}(r) = \frac{1}{4r} \int_0^{r} \tau_{00}(r')\, r'^2  dr' +   \frac{1}{4} \int_r^{\infty} \tau_{00}(r') \, r' dr'.
\end{equation}
Inserting the coefficients (\ref{tau_decomp_coef}) and carrying out the integrations then yields
\begin{subequations} \label{u_result}
\begin{align}
u_{00}(r) & = -\frac{\sqrt{\pi }}{6 \sigma^2}  \Ab e^{-(r / \sigma)^2} \left(2 r^2+\sigma ^2\right) \\
u_{20}(r) & = -\sqrt{\frac{\pi}{5}}\frac{\Ab}{24 r^3
   \sigma ^2}  \Big( 3 \sqrt{\pi } \sigma ^5 \erf \left(\frac{r}{\sigma }\right) -\\ 
   & ~~~~~~~~~~ 2 r e^{-( r/\sigma)^2} \left(2 r^2 \sigma ^2+4 r^4+3 \sigma ^4\right)\Big) \nonumber 
\end{align}
\end{subequations}
where the error function $\erf(z)$ is defined as
\begin{equation}
    \erf(z) \equiv \frac{2}{\sqrt{\pi}} \int_0^z e^{-t^2} dt.
\end{equation}
The leading-order terms in a Taylor expansion of $\erf(z)$ about $z = 0$ are given by
\begin{equation}
    \erf(z) = \frac{2}{\sqrt{\pi}} \left( z - \frac{z^3}{3} \right) + \mathcal{O}(z^5),
\end{equation}
so that $u_{20}(r)$ remains finite as $r \rightarrow 0$.

Finally, we assemble the spatial metric by inserting the expressions (\ref{q_decomp}) and (\ref{u_decomp}) into the line element (\ref{metric_lin}), which, together with the assumption of time symmetry, completes the construction of (linear) Brill wave initial data $\gamma^{\rm B}_{ij}$.  As for the Teukolsky waves, we then define the Brill wave perturbations from
\begin{equation} \label{eq:h_brill}
h^{\rm B}_{ij} = \gamma^{\rm B}_{ij} - \eta_{ij}.
\end{equation}

\section{Comparisons}
\label{sec:compare}

Superficially, the Teukolsky wave initial data $\gamma_{ij}^{\rm T}$ of Section \ref{sec:indata:teukolsky} and the Brill wave data $\gamma^{\rm B}_{ij}$ of Section \ref{sec:indata:brill} appear different; for example, the $(r\theta)$ component of the spatial metric vanishes for Brill data, $\gamma^{\rm B}_{r\theta} = 0$, but does not for Teukolsky data.  Such a direct comparison is not meaningful, however, because the data appear in different gauges.  We therefore adopt three different approaches to make such a comparison: in Section \ref{sec:transform} we transform the Brill data directly into the transverse-traceless (TT) gauge of the Teukolsky data, in Section \ref{sec:moncrief} we employ the gauge-invariant Moncrief formalism, and finally, in Section \ref{sec:kretschmann}, we construct and compare the Kretschmann scalar for both sets of data.

\subsection{Gauge transformations}
\label{sec:transform}

The Teukolsky data of Section \ref{sec:indata:teukolsky} adopt TT gauge, which, for the purely spatial metric perturbation $h^{\rm T}_{ij}$, means that
\begin{subequations} \label{TT}
\begin{align}
    \eta^{ij} h_{ij} & = 0,  \label{TT1} \\
    \partial^j h_{ij} & = 0  \label{TT2}
\end{align}
\end{subequations}
({\it cf.}~eqs.~(2) and (3) in \cite{Teu82}).  Note that we will adopt Cartesian coordinates in this Section, so that $\eta_{ij} = \mbox{diag}(1,1,1)$ and covariant derivatives associated with $\eta_{ij}$ become partial derivatives.  Note also that both the Teukolsky and the Brill data satisfy the linearized vacuum Hamiltonian constraint
\begin{align} \label{ham}
\partial^i\partial^j h_{ij}-\partial^i\partial_i h^k{}_{k}=0\,.
\end{align}
Our goal is now to transform the Brill data of Section \ref{sec:indata:brill} into TT  gauge.

The linearized spatial gauge freedom is generated by a spatial one-form $\xi_i$ and can be expressed as
\begin{align}
    h^{\rm B'}_{ij}=h^{\rm B}_{ij}-2\partial_{(i}\xi_{j)},
    \label{eqn:linear_gauge}
\end{align}
where $h^{\rm B'}_{ij}$ represents the Brill wave perturbations in the new gauge.  Applying the condition (\ref{TT2}) to this new gauge we obtain
\begin{equation} 
   \partial^j \partial_j \xi_i + \partial_i \partial^j \xi_j = \partial^j h_{ij}^{\rm B}.
\end{equation}
We may solve this equation by decomposing $\xi_i$ according to
\begin{align}
\xi_i=\hat{\xi}_i+\partial_i\varphi\,,\label{eqn:xi_decomp}
\end{align}
with
\begin{subequations} \label{TT_xi}
\begin{align}
\partial^i\partial_i\varphi&= \frac{1}{2} h^{\rm B}\,, \label{eqn:xi_varphi} \\
\partial^j\partial_j\hat{\xi}_i&=\partial^j h_{ij}^{\rm B} -\partial_i h^{\rm B} \label{eqn:xihat_i}
\end{align}
\end{subequations}
where $h^{\rm B} \equiv \eta^{ij} h^{\rm B}_{ij}$.  

Taking the divergence of~\eqref{eqn:xihat_i} we see that
\begin{align}
\partial^j\partial_j(\partial^i\hat{\xi}_i)&=\partial^i\partial^jh_{ij}-\partial^i\partial_i h^k{}_k=0\,,
\end{align}
where we have used the Hamiltonian constraint (\ref{ham}) in the last step.  Given suitable boundary conditions, this implies that the divergence of $\hat \xi_i$ vanishes everywhere, 
\begin{equation} \label{hatxidiv}
\partial^i \hat \xi_i = 0.
\end{equation} 
so that the decomposition~\eqref{eqn:xi_decomp} splits the generator $\xi_i$ into transverse and longitudinal parts.  Finally, we may take the trace of (\ref{eqn:linear_gauge}) to see that
\begin{equation}
    h^{\rm B'} = h^{\rm B} - 2 \partial^i \xi_i = h^{\rm B} - 2 \, \partial^i \partial_i \varphi = 0,
\end{equation}
where we have used (\ref{hatxidiv}) in the second equality and (\ref{eqn:xi_varphi}) in the third.  This shows that, with the decomposition (\ref{eqn:xi_decomp}) of $\xi_i$ satisfying equations (\ref{TT_xi}), the new metric (\ref{eqn:linear_gauge}) will indeed satisfy both TT conditions (\ref{TT}).

\begin{figure}[t]
    \includegraphics[width = 0.45 \textwidth]{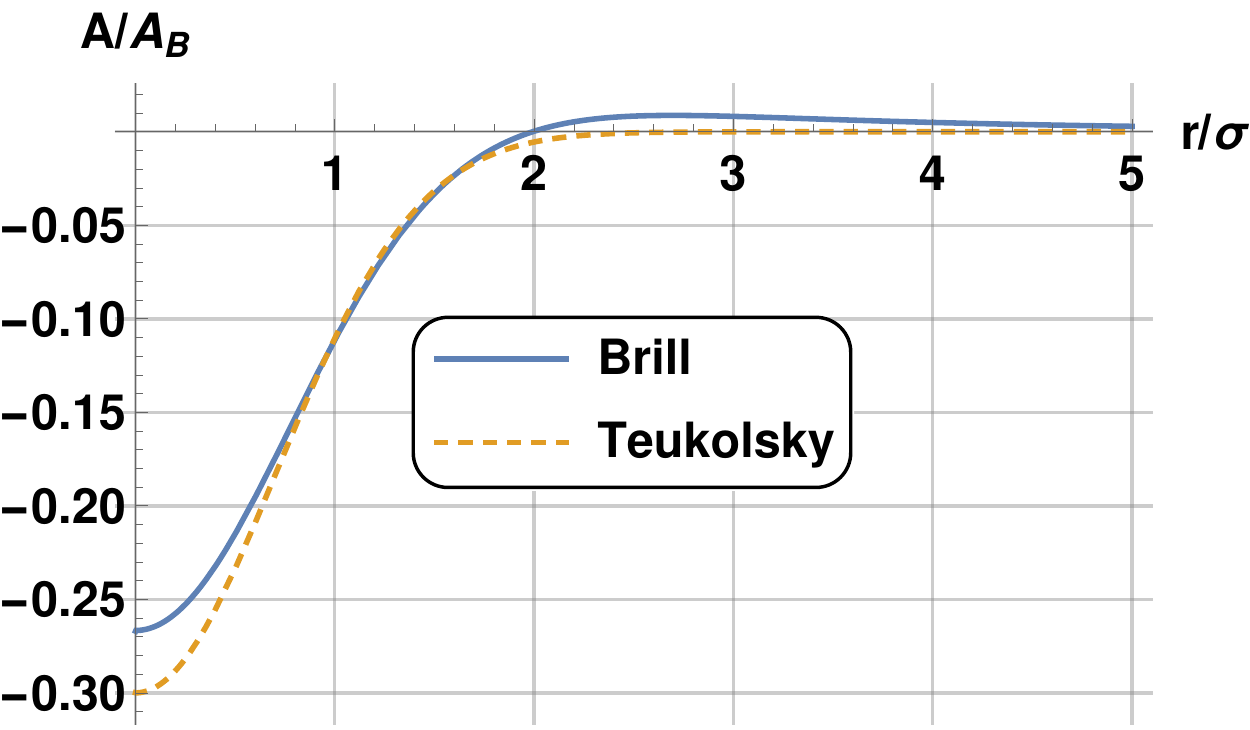}
    
    \includegraphics[width = 0.45 \textwidth]{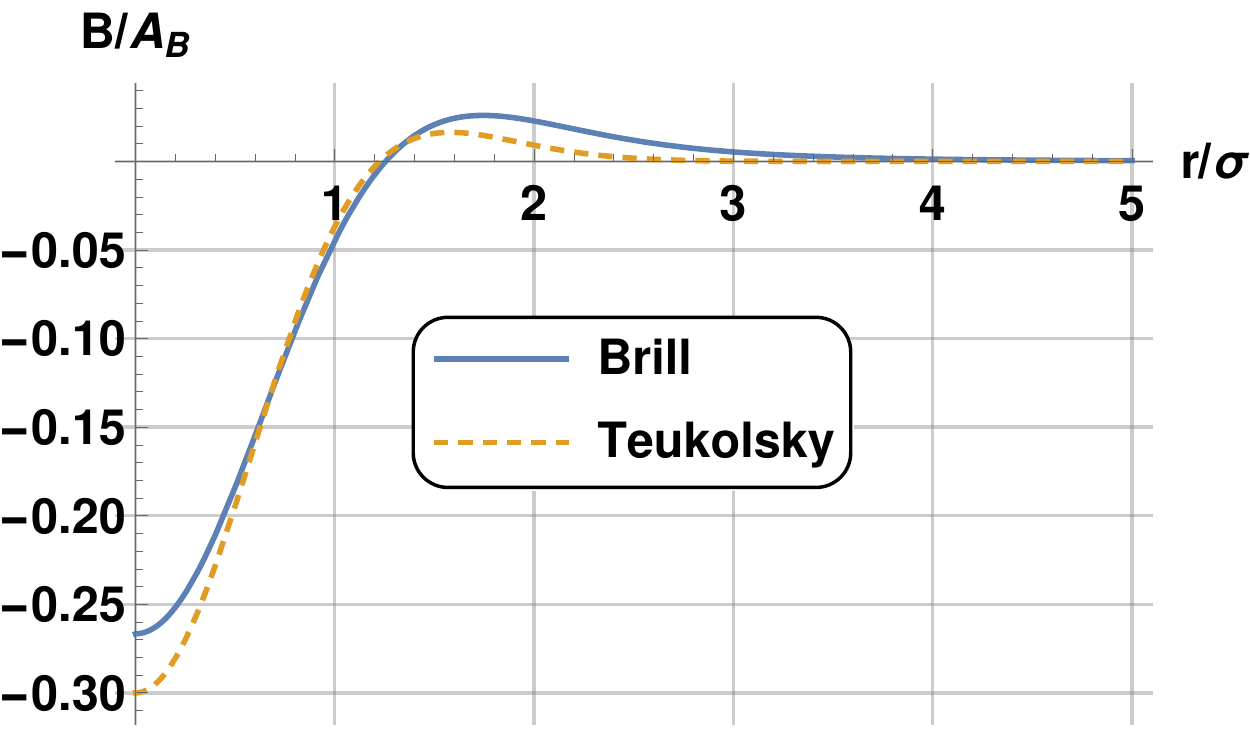} 
        
    \includegraphics[width = 0.45 \textwidth]{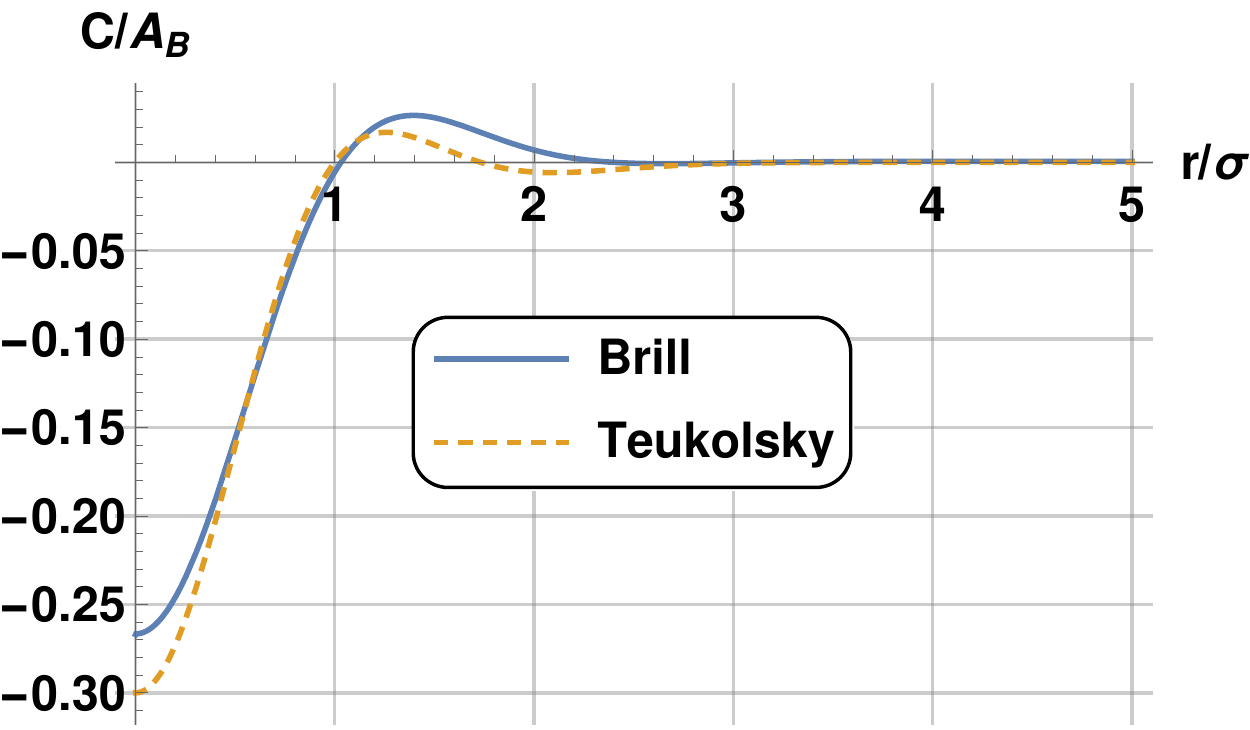}
    
    \caption{Comparisons of the functions $A$ (top panel), $B$ (middle panel, and $C$ (bottom panel) for Teukolsky waves and Brill waves, both expressed in TT gauge.  From these functions, which are listed in eqs.~(\ref{eq:A_coef_Teuk}) and (\ref{eq:A_coef_Brill}), respectively, the initial spatial metric $\gamma_{ij}$ can be computed from (\ref{eq:A_h_gen}).  For the purposes of these comparisons we adopt $\At = \Ab / 80$ and $\sigma = \lambda$, and show the functions $A$, $B$, and $C$ divided by $\Ab$ for both sets of data.}
    \label{fig:1}
\end{figure}

We invert the Laplace operators in (\ref{TT_xi}) using the same approach as in Section \ref{sec:indata:brill}.   We solve equation (\ref{eqn:xihat_i}) for Cartesian components of $\hat \xi_i$, but carry out the integration over the Green function using spherical polar coordinates together with the expansion (\ref{green_expand}).  Once $\hat \xi_i$ and $\varphi$ have been found, we assemble $\xi_i$ from (\ref{eqn:xi_decomp}) and compute the Brill initial data in TT gauge from (\ref{eqn:linear_gauge}).  Quite remarkably, after carrying out the transformation to TT gauge, the Brill initial data for the seed function (\ref{seed1}) can also be expressed in the form (\ref{eq:A_h_gen}) of a Teukolsky wave, but now with the coefficients $A^{\rm B}$, $B^{\rm B}$, and $C^{\rm B}$, given by the expressions (\ref{eq:A_coef_Brill}).  

In Fig.~\ref{fig:1} we compare these coefficients with those for the Teukolsky perturbations.  We see that, with a suitable rescaling of the amplitudes, $\At = \Ab/80$, the qualitative features of the two sets of initial data, for the given seed functions, indeed appear quite similar.

We note, however, that the transformation of the linearized Brill wave initial data to TT gauge results in a purely quadrupolar Teukolsky wave only for the specific angular dependence of the seed function (\ref{seed1}).  In general, linearized Brill wave initial data are superpositions of waves with different multipole moments, as one might have suspected, but for the seed function (\ref{seed1}) all multipoles different from the quadrupole moment are suppressed.  We will explore this result in more detail below.

\subsection{Gauge-invariant Moncrief formalism}
\label{sec:moncrief}

As a second approach to comparing the Teukolsky and Brill data we employ the gauge-invariant Moncrief formalism (see \cite{Mon74}; see also \cite{NagR05} for a review as well as Section 9.4.1 in \cite{ShaB10} for a textbook treatment).  In general, the Moncrief formalism assumes that the spacetime metric can be decomposed into a background metric $g^B_{ab}$ given by the Schwarzschild metric and a perturbation $h_{ab}$.  In our specific case the background metric is flat, and hence corresponds to a zero-mass Schwarzschild spacetime.  The perturbation $h_{ab}$ is then decomposed into scalar, vector, and tensor spherical harmonics of even or odd parity, from which the gauge-invariant Moncrief functions $R_{\ell m}$ can be computed for each mode $\ell$ and $m$.  

For both the Teukolsky and the Brill data, only even-parity contributions enter the decomposition of the perturbative metric, for which we may follow the prescription starting with eq.~(9.77) in \cite{ShaB10}.  Specifically, we compute, for both the Teukolsky data $\gamma^{\rm T}_{ij}$ and the Brill data $\gamma^{\rm B}_{ij}$, the projections $H_{2 \ell m}$, $h_{1\ell m}$, $K_{\ell m}$, and $G_{\ell m}$ from eqs.~(9.78) through (9.81).  In these integrals, the components of the tensor spherical harmonics can be expressed in terms of functions $W_{\ell m}$ and $X_{\ell m}$, which we list in Appendix \ref{sec:moncrief:ang_fcts}.  For example, we compute $G_{\ell m}$ from 
\begin{equation} \label{eq:Glm}
    G_{\ell m} = \frac{1}{2 (\ell -1) \ell (\ell + 1) (\ell + 2)}
    \frac{1}{r^2} \int \gamma_- W^*_{\ell m} \, d \Omega
\end{equation}
(where $\gamma_- \equiv \gamma_{\theta\theta} - \gamma_{\phi\phi} / \sin^2 \theta$, and where we have assumed $\gamma_{\theta\phi} = 0$).

In the next step, we find the functions $k_{1\ell m}$ and $k_{2\ell m}$ from (9.88) and (9.89) in \cite{ShaB10}.  Finally, these functions can be combined into the gauge-invariant Moncrief functions $R_{\ell m}$ as in (9.87) in \cite{ShaB10}.  

For the Teukolsky waves of Section \ref{sec:indata:teukolsky}, we list all intermediate results in Appendix \ref{sec:moncrief:teukolsky}.  Since these data are constructed as an axisymmetric, purely quadrupolar wave, it is not surprising that the only non-vanishing terms are those with $\ell = 2$ and $m = 0$. The final result for the gauge-invariant Moncrief function $R^{\rm T}_{20}$ is
\begin{align}
    R^{\rm T}_{20} & = -\sqrt{\frac{\pi }{5}} \frac{8 \At }{\lambda ^4}r^3 e^{-(r/ \lambda)^2} \left(2 r^2-7 \lambda ^2\right)
    \label{eq:R20_T}
\end{align}

\begin{figure}[t]
    \centering
    \includegraphics[width = 0.45 \textwidth]{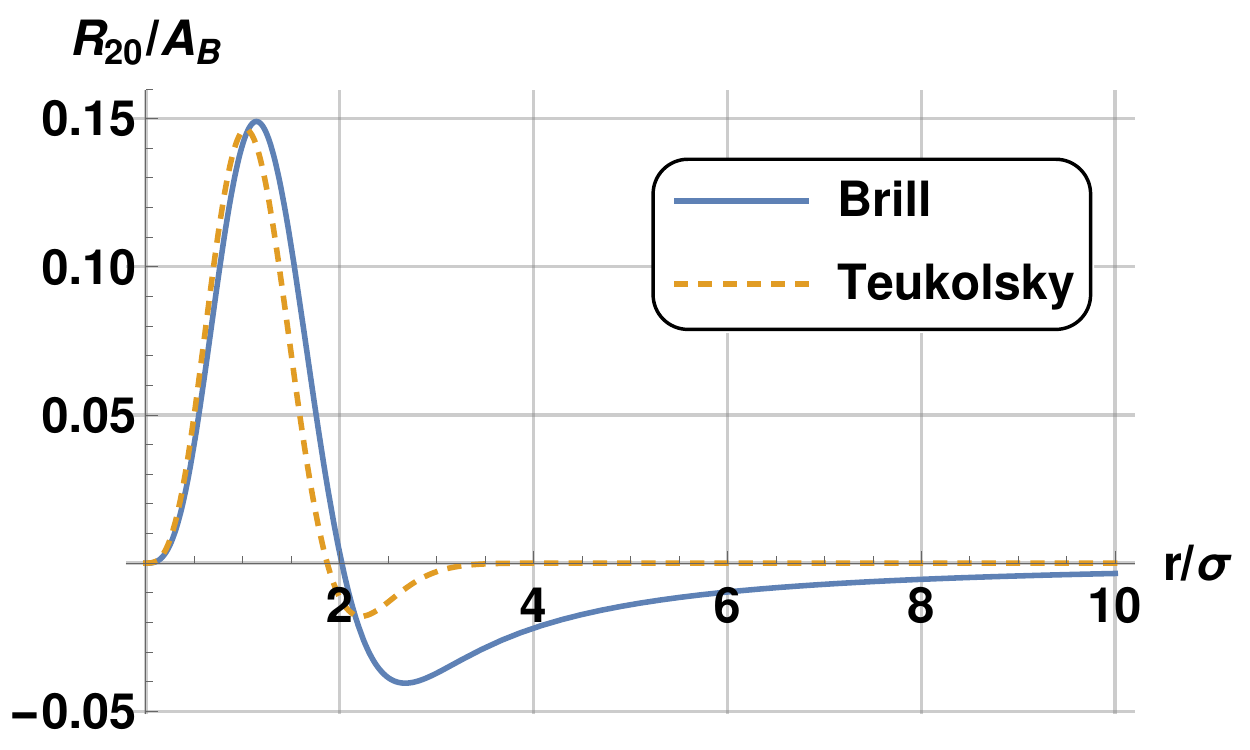}
    \caption{The gauge-invariant Moncrief function $R_{20} / \Ab$ for Teukolsky (orange dashed line) and Brill (blue continuous line) waves.  As in Fig.~\ref{fig:1} we choose $\At = \Ab / 80$ and $\lambda = \sigma$ and plot both functions in units of the amplitude $\Ab$.}
    \label{fig:2}
\end{figure}

The Brill waves of Section \ref{sec:indata:brill}, on the other hand, are not purely quadrupolar by construction.  In Section \ref{sec:transform} we have already seen that, for the special choice of the seed function (\ref{seed1}), a transformation of the data to TT gauge again results in a purely quadrupolar Teukolsky wave.  It is therefore not surprising that, in this case, the only non-vanishing Moncrief function is again that with $\ell = 2$ and $m = 0$.  Alternatively, we may apply the Moncrief formalism to the Brill wave in its original gauge of Section \ref{sec:indata:brill}.  In this case, the intermediate results for the projections $H_{2 \ell m}$, $h_{1\ell m}$, $K_{\ell m}$, and $G_{\ell m}$ as well as the functions $k_{1\ell m}$ and $k_{2\ell m}$ are listed in Appendix \ref{sec:moncrief:brill}.  The Moncrief function $R_{20}$ is, by construction, independent of gauge, and given by
\begin{align}
    R^{\rm B}_{20} & =\sqrt{\frac{\pi}{5} } \Ab \left[ \frac{1   }{6 r \sigma ^2} e^{-(r/\sigma) ^2} \left(4 r^4 + 2 r^2 \sigma ^2+3 \sigma ^4\right)\right. - \nonumber \\
    & ~~~~~~~~~~~~~~ \frac{ \sqrt{\pi} }{4} \left. \frac{\sigma^3}{r^2}\erf\left(\frac{r}{\sigma }\right) \right]
    \label{eq:R20_B}
\end{align}
Both (\ref{eq:R20_T}) and (\ref{eq:R20_B}) can also be written in the form
\begin{equation}
    R_{20} = \frac{r}{6} \sqrt{ \frac{\pi}{5} } \left(
    r \partial_r A - 6 A - 6 B + 12 C \right),
\end{equation}
with the functions $A$, $B$, and $C$ given by (\ref{eq:A_coef_Teuk}) for Teukolsky waves, and by (\ref{eq:A_coef_Brill}) for Brill waves
(see also exercise 9.7 in \cite{ShaB10}).  

In Fig.~\ref{fig:2} we compare the Moncrief functions (\ref{eq:R20_T}) and (\ref{eq:R20_B}).  While the two results for Teukolsky and Brill wave evidently differ quantitatively, their general qualitative features are, in fact, quite similar -- which is consistent with our findings of Section \ref{sec:transform}.

Finally, it is instructive to consider multipole moments with $\ell > 2$ for the Brill wave initial data.  Starting with these data in the gauge of Section \ref{sec:indata:brill}, the projections $H_{2 \ell m}$, $h_{1\ell m}$, and $K_{\ell m}$ must all vanish identically for $\ell > 2$, but $G_{\ell m}$, given by (\ref{eq:Glm}), could be nonzero.  To evaluate this term for Brill waves we observe that, from (\ref{metric_lin}), we have $\gamma_- = 2 q$, which contains both monopole and quadrupole terms (see eq.~\ref{q_decomp}).  We also note that the functions $W_{\ell 0}$ can be written as a linear combination of spherical harmonics $Y_{\ell' 0}$ with $\ell' \leq \ell$ (see Appendix \ref{app:wlm}).  Using (\ref{eq:wl0_exp}), the integral in (\ref{eq:Glm}) may therefore be written as
\begin{align} \label{eq:Wlm_integral}
    \frac{1}{r^2} & \int \gamma_- W^*_{\ell 0} \, d \Omega =
    4 \sqrt{2 \ell + 1} \int q ( \sqrt{5} Y_{20} + Y_{00} ) \, d \Omega \nonumber \\
    & = 4 \sqrt{2 \ell + 1} \left( \sqrt{5} \, q_{20} + q_{00} \right), ~~~~~~~(\ell > 2~\mbox{even})
\end{align}
where we have employed the decomposition (\ref{q_decomp}) in the last step.  In general, this integral will therefore {\em not} vanish, and will instead give rise to multipole moments higher in order than $\ell = 2$.  For the seed function (\ref{seed1}), however, we have $q_{20} = - q_{00} / \sqrt{5}$ (see eqs.~\ref{q_decomp_coef}), leading to an exact cancellation in (\ref{eq:Wlm_integral}), and therefore to a vanishing of all higher-order multipole moments.  This result is consistent with our finding in Section \ref{sec:transform} that, when transformed to TT gauge, Brill waves become purely quadrupolar if the seed function has the angular dependence of (\ref{seed1}).

\subsection{Kretschmann scalar}
\label{sec:kretschmann}

\begin{figure*}[t]
\centering
\subfigure{\includegraphics[width=0.45 \textwidth]{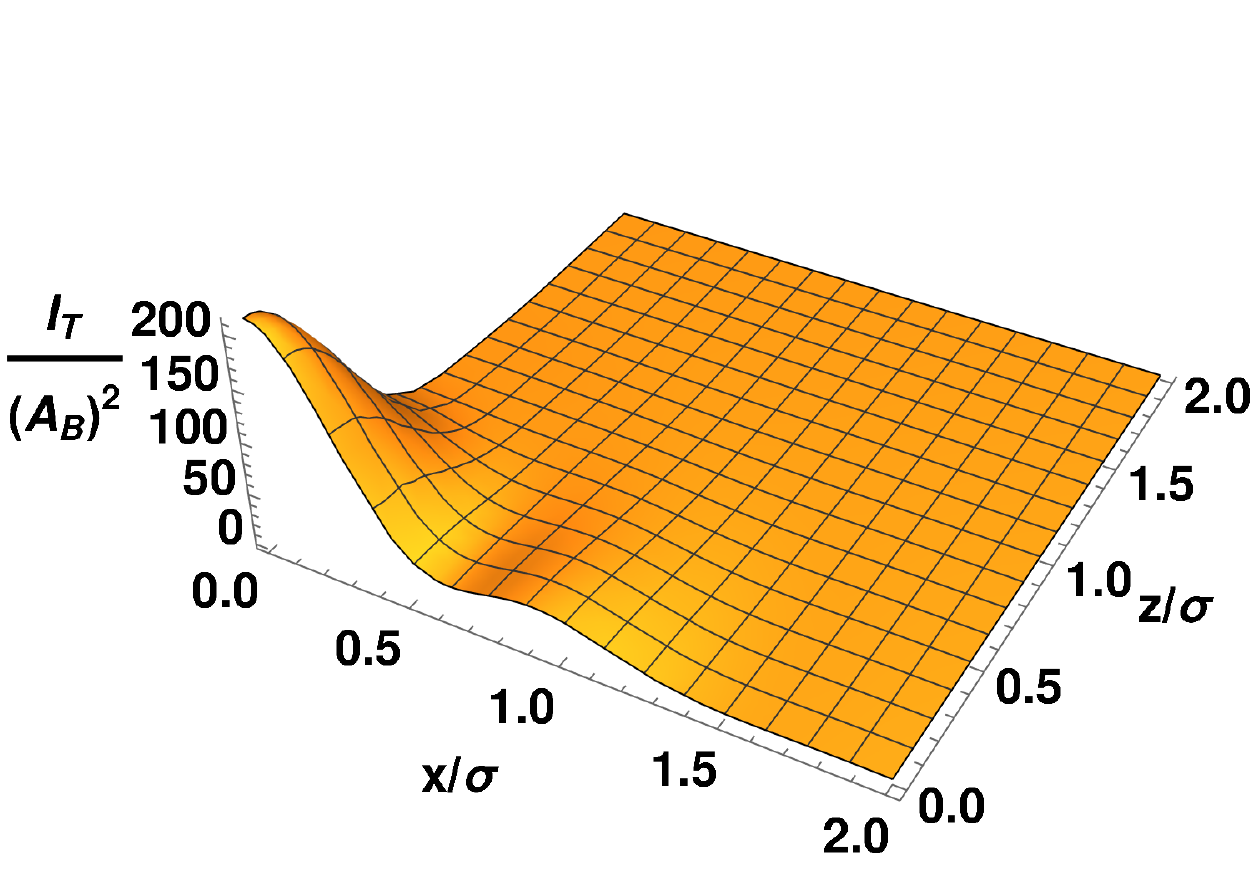}}
\subfigure{\includegraphics[width=0.45 \textwidth]{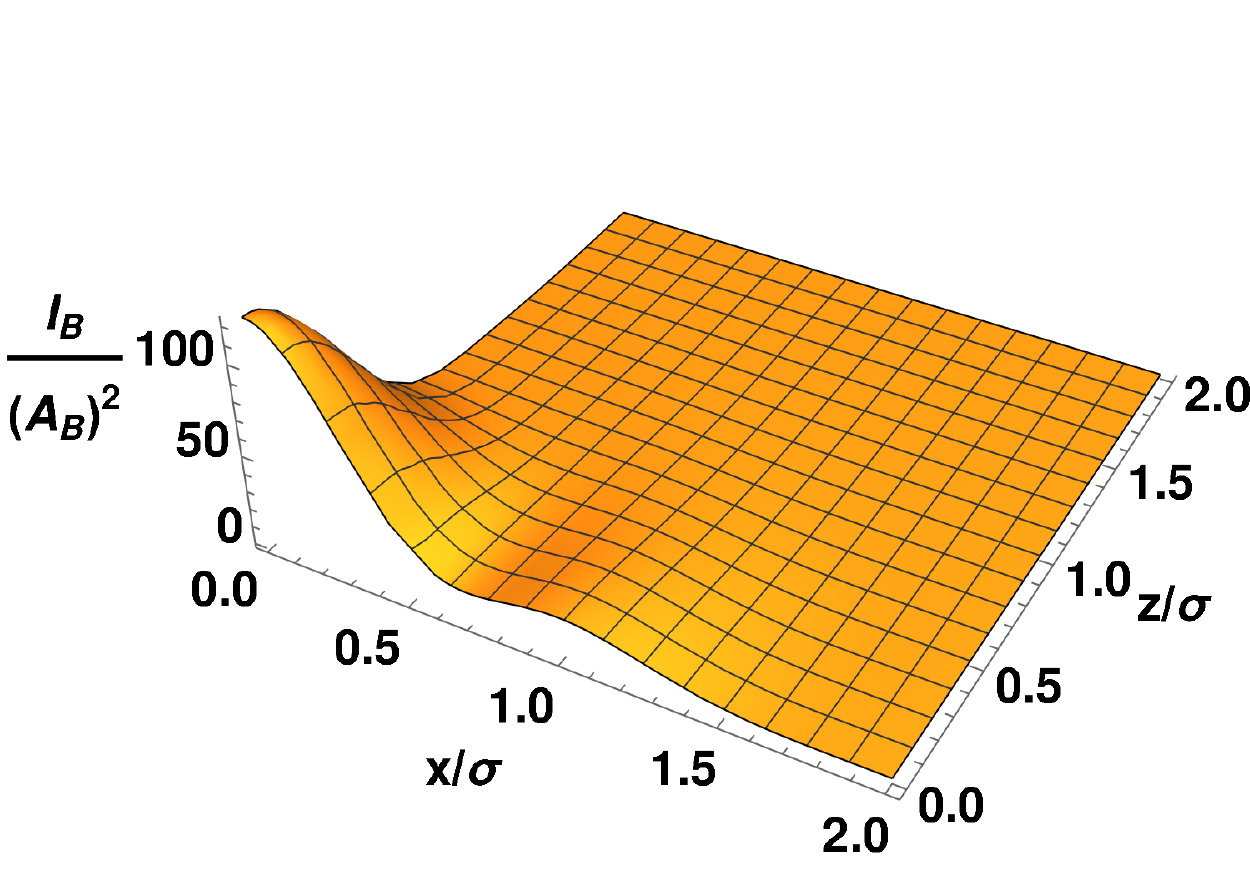}}
\caption{The Kretschmann scalar $I$ for both Teukolsky waves (left panel) and Brill waves (right panel). As in Figs.~\ref{fig:1} and \ref{fig:2} we adopt $\At = \Ab / 80$ and $\lambda = \sigma$, and show $I$ divided by $\Ab^2$ for both waves.} 
\label{fig:3}
\end{figure*}

\begin{figure}[t]
    \centering
    \includegraphics[width = 0.45 \textwidth]{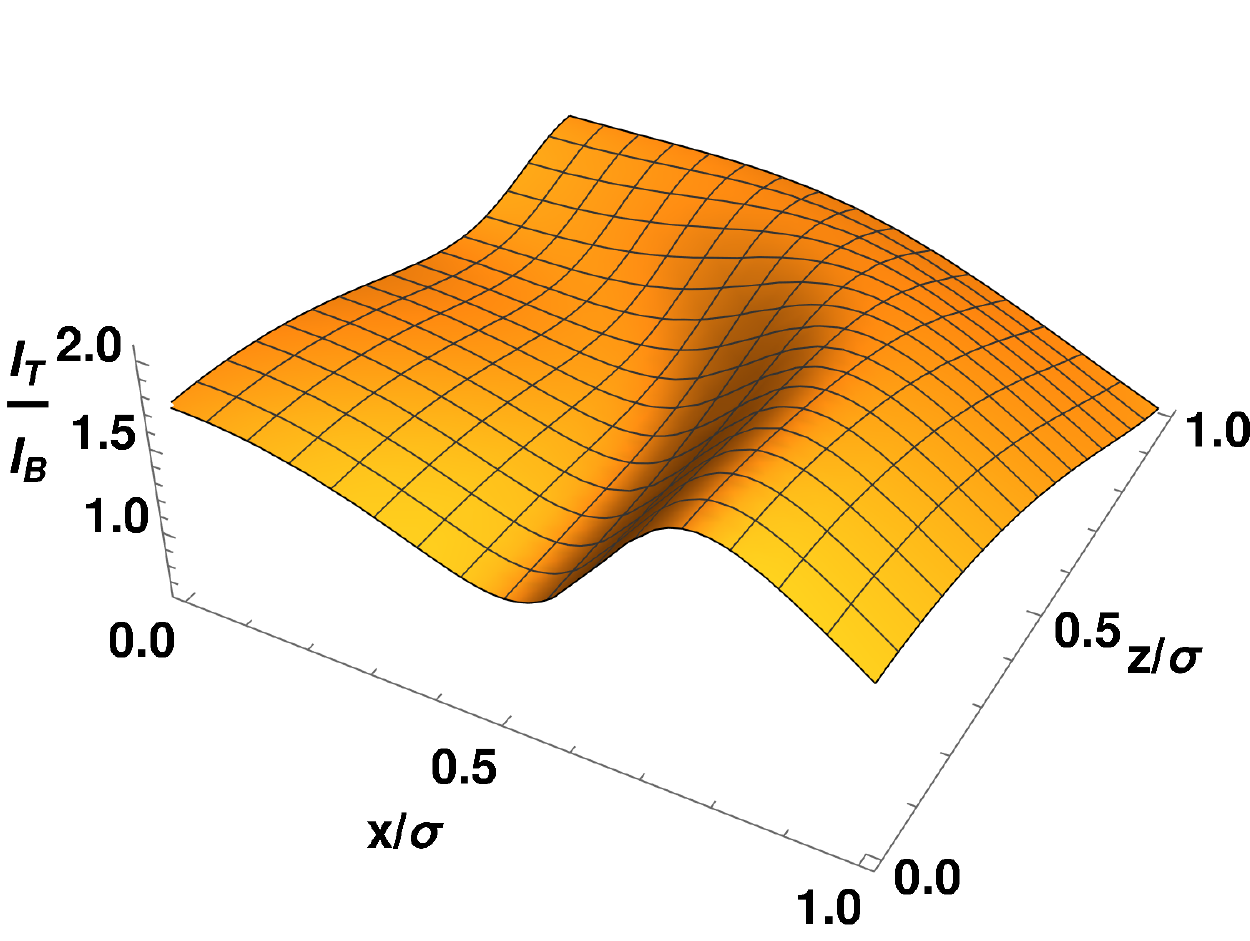}
    \caption{The ratio $I_{\rm T} / I_{\rm B}$ between the Kretschmann scalars for Teukolsky and Brill waves. As in the previous figures we adopt $\At = \Ab / 80$ and $\lambda = \sigma$.}
    \label{fig:4}
\end{figure}

As a third way of comparing the Teukolsky and Brill data we compute the Kretschmann scalar
\begin{equation}
    I = {}^{(4)}R^{abcd} \, {}^{(4)}R_{abcd}, 
    \label{eq:Kretschmann}
\end{equation}
where ${}^{(4)}R_{abcd}$ is the (four-dimensional) spacetime Riemann tensor of the spacetime.  For our time-symmetric vacuum data, the Kretschmann scalar can be expressed in terms of the (three-dimensional) spatial Ricci tensor $R_{ij}$ only,
\begin{equation}
I  = 8 \, \gamma^{ij} \gamma^{kl} R_{ik} R_{jl}.
\label{eq:K_lin}
\end{equation}
We compute the Kretschmann scalar $I$ for both the Teukolsky data of Section \ref{sec:indata:teukolsky} and the Brill data of Section \ref{sec:indata:brill}, and compare the results in Fig.~\ref{fig:3}.  As in our previous comparisons, we see that all qualitative features are very similar.

As a more direct way of comparing the Kretschmann scalars $I_{\rm T}$ and $I_{\rm B}$ we show their ratio in Fig.~\ref{fig:4}.  Evidently, this ratio is defined only up to a constant related to the ratio between the two amplitudes $\At$ and $\Ab$; as in the previous figures we fix this ratio by adopting $\At = \Ab / 80$ in Fig.~\ref{fig:4}.  Given that the ratio shows some spatial variations, we see that the Kretschmann scalars $I_{\rm T}$ and $I_{\rm B}$ are indeed different quantitatively, despite the similarity in their qualitative features.

\section{Summary and discussion}
\label{sec:summary}

We compared, at the linear level, two common approaches that have been adopted in the construction of gravitational wave initial data,  namely Teukolsky data \cite{Teu82} (see Section \ref{sec:indata:teukolsky}) and Brill data \cite{Bri59} (see Section \ref{sec:indata:brill}).  Both approaches employ a seed function, for which we chose, following numerous other authors, the Gaussian profiles (\ref{eqn:Teuk_Seed}) and (\ref{seed1}).  While the Teukolsky waves are constructed as purely quadrupolar $\ell = 2$ waves, the Brill waves are not.  

Since the two sets of initial data appear in different spatial gauges, they cannot be compared directly.  Instead we adopted three different approaches to compare the data: we transformed the Brill data into the TT gauge of the Teukolsky data (Section \ref{sec:transform}), computed the gauge-invariant Moncrief functions (Section \ref{sec:moncrief}), and evaluated the Kretschmann scalar (Section \ref{sec:kretschmann}).  

To our surprise we found that, while linearized Brill waves will in general {\em not} be purely quadrupolar, and will instead be superpositions of waves with different multipole moments, for special seed functions with the angular dependence of (\ref{seed1}) all higher-order moments cancel out exactly, casting the Brill waves again as a purely quadrupolar wave.  While these waves are not identical to Teukolsky waves with the seed function (\ref{eqn:Teuk_Seed}), they share many qualitative features in all our comparisons.

Teukolsky and Brill wave data play an important role in the context of vacuum critical collapse, where they have been adopted by a number of different authors.  Our study was motivated by the observations that (a) the two types of data appear to lead to different threshold solutions (with different critical exponents $\gamma$; see \cite{LedK21}), and also appear to behave different numerically (see also \cite{Hiletal13}), and (b) initial data with different multipole moments lead to quantitatively different threshold solutions in the critical collapse of electromagnetic waves (see \cite{BauGH19,PerB21}).  The latter suggests that higher-order multipole moments present in the Brill data might result in the observed differences in their evolution from those of Teukolsky data.  However, as we discussed above, for precisely the seed function typically employed for Brill waves those higher-order multipole moments vanish exactly.  We therefore conclude that the multipole structure of Brill waves cannot be held responsible for the observed differences.

We note, however, that even for our choices of the seed functions the data are not identical.  We have discussed before that they appear in different gauges; moreover, while a Brill wave with the seed function (\ref{seed1}) is quadrupolar, it corresponds to a seed function that is different from (\ref{eqn:Teuk_Seed}).  While it may well be worth exploring whether either one of these differences is related to the observed differences in the evolution of the data, it is also possible that the latter are related to nonlinear effects, which we have ignored in our analysis here.

\acknowledgments

The work was supported in parts by the FCT (Portugal) IF Programs IF/00577/2015 and PTDC/MAT-APL/30043/2017, the IDPASC program PD/BD/135434/2017 and Project No.\ UIDB/00099/2020, as well as National Science Foundation (NSF) grant PHY-2010394 to Bowdoin College.

\begin{appendix}

\section{Expressions for the spatial metric in TT gauge}
\label{app:metric}

When expressed in TT gauge, the non-vanishing initial metric perturbations for the Teukolsky data of Section \ref{sec:indata:teukolsky} and Brill data of Section \ref{sec:indata:brill}, expressed in spherical polar coordinates, can be written in the form
\begin{subequations} \label{eq:A_h_gen}
\begin{align}
    h_{rr} & = A f_{rr}, \\ 
    h_{r\theta} & = r B f_{r\theta}, \\
    h_{\theta\theta} & = r^2 ( C f_{\theta\theta}^{(1)} + A f_{\theta\theta}^{(2)} ), \\
    h_{\phi\phi} & = r^2 \sin^2 \theta ( C f_{\phi\phi}^{(1)} + A f_{\phi\phi}^{(2)} ).
\end{align}
\end{subequations}
In (\ref{eq:A_h_gen}), the angular functions are those for $\ell = 2$ and $m = 0$,

\begin{subequations} \label{eq:A_ang_fcts}
\begin{align}
    f_{rr} & = 2 - 3\sin^2(\theta) \\
    f_{r \theta} & = -3\sin(\theta) \cos(\theta) \\
    f^{(1)}_{\theta \theta} & = 3 \sin^2(\theta) \\
    f^{(2)}_{\theta \theta} & = -1 \\
    f^{(1)}_{\phi \phi} & = - f^{(1)}_{\theta \theta} \\ 
    f^{(2)}_{\phi \phi} & = 3 \sin^2 (\theta) -1.
    \label{eq:angular_funcs}
\end{align}
\end{subequations}
For Teukolsky data, with the seed function $F(t,r)$ given by (\ref{eqn:Teuk_Seed}), the coefficients $A$, $B$, and $C$, evaluated at the moment of time symmetry $t = 0$, are given by

\begin{subequations} \label{eq:A_coef_Teuk}
\begin{align}
    A^{\rm T} & = -24 \At e^{- (r/\lambda)^2}, \\
    B^{\rm T} & = \frac{8 \At}{\lambda^2} e^{- (r/\lambda)^2}(2r^2 - 3\lambda^2),  \\
    C^{\rm T} & = \frac{8 \At}{\lambda^4} e^{- (r/\lambda)^2}(r^4 - 4r^2\lambda^2 + 3\lambda^4). 
\end{align}
\end{subequations}
For Brill data, once transformed to TT gauge as described in Section \ref{sec:transform}, these coefficients take the form 
\begin{subequations} \label{eq:A_coef_Brill}
\begin{align}
    A^{\rm B} & = \frac{\Ab \sigma ^2}{8 r^5} \Big[2 r e^{-(r/\sigma)^2} \left(4 r^2+9 \sigma ^2\right) + \nonumber \\
    & ~~~~~~~~~~   \sqrt{\pi } \sigma  \left(2 r^2-9 \sigma ^2\right) \erf \left(\frac{r}{\sigma }\right)\Big], \\
    B^{\rm B} & = -\frac{\Ab}{12 r^5} \Big[ 2 r e^{-(r/\sigma)^2} \left(4 r^4+6 r^2 \sigma ^2+9 \sigma ^4\right)- \nonumber  \\
    & ~~~~~~~~~~   9 \sqrt{\pi } \sigma ^5 \erf \left(\frac{r}{\sigma }\right)\Big],  \\
    C^{\rm B} & = \frac{\Ab}{96 r^5 \sigma ^2} \left[ 2 r e^{-\frac{r^2}{\sigma ^2}} \left(16 r^6+36 r^2 \sigma ^4+63 \sigma^6\right) \right. +\nonumber \\ 
    & ~~~~~~~~~~ \left.  3 \sqrt{\pi } \sigma ^5 \left(2 r^2-21 \sigma ^2\right) \erf \left(\frac{r}{\sigma }\right) \right]. 
\end{align}
\end{subequations}

\section{Construction of Moncrief functions}
\label{app:moncrief}

\subsection{Auxiliary angular functions}
\label{sec:moncrief:ang_fcts}

In the construction of the gauge-invariant Moncrief functions it is useful to express the components of the tensor spherical harmonics in terms of the functions
\begin{subequations}
\begin{align}
    W_{\ell m} & = \left(\partial^2_\theta - \cot \theta \, \partial_\theta - \frac{1}{\sin^2 \theta} \partial^2 \phi \right) 
    Y_{\ell m} \\
    X_{\ell m} & = 2 \partial_\phi \left( \partial_\theta - \cot \theta \right) Y_{\ell m}
\end{align}
\end{subequations}
(see, e.g., Section 9.4.1 and Appendix D in \cite{ShaB10}).  For $\ell = 2$ and $m = 0$, these functions reduce to
\begin{subequations}
\begin{align}
    W_{20} & = \frac{3}{2} \sqrt{\frac{5}{\pi }} \sin ^2(\theta ) \\
    X_{20} & = 0
\end{align}
\end{subequations}

\subsection{Teukolsky waves}
\label{sec:moncrief:teukolsky}

As one might expect for an axisymmetric, purely quadrupolar wave, the only non-vanishing terms for the Teukolsky wave of Section \ref{sec:indata:teukolsky} are those with 
$\ell = 2$ and $m = 0$.  From eqs.~(9.78) through (9.81) in \cite{ShaB10} we compute the functions $H_{220}$, $h_{120}$, $K_{20}$, and $G_{20}$ to be
\begin{subequations}
\begin{align}
    H^{\rm T}_{220} & = -96 \sqrt{\frac{\pi }{5}} \At e^{- (r/\lambda)^2}  \\
    h^{\rm T}_{120} & = \sqrt{\frac{\pi }{5}} \frac{16  \At }{\lambda ^2}r e^{- (r/\lambda)^2}  \left(2 r^2-3 \lambda ^2\right) \\
    K^{\rm T}_{20} & = 48 \sqrt{\frac{\pi }{5}} \At e^{- (r/\lambda)^2}  \\
    G^{\rm T}_{20} & = -\sqrt{\frac{\pi }{5}}\frac{8  \At }{\lambda ^4}e^{- (r/\lambda)^2}  \left(2 r^4-8 \lambda ^2 r^2+3 \lambda ^4\right).
\end{align}
\end{subequations}
Following (9.88) and (9.89) we can then combine these functions to form
\begin{subequations}
\begin{align}
    k^{\rm T}_{120} & = \sqrt{\frac{\pi }{5}} \frac{32 \At r^2 }{\lambda ^6}e^{- (r/\lambda)^2}  \left(2 r^4-15 \lambda ^2 r^2+21 \lambda ^4\right) \\
    k^{\rm T}_{220} & = - \sqrt{\frac{\pi }{5}} \frac{48 \At r^2}{\lambda ^6} e^{- (r/\lambda)^2} \left(2 r^4-13 \lambda ^2 r^2 + 14 \lambda ^4\right).
\end{align}
\end{subequations}
Finally, the gauge-invariant Moncrief function $R_{20}$ for the Teukolsky wave of Section \ref{sec:indata:teukolsky}, computed from (9.87) in \cite{ShaB10}, is given by
\begin{align}
    R^{\rm T}_{20} & = -\sqrt{\frac{\pi }{5}}\frac{8  \At r^3}{\lambda ^4} e^{- (r/\lambda)^2}  \left(2 r^2-7 \lambda ^2\right)
\end{align}
(see also exercise 9.7 in \cite{ShaB10}).

\subsection{Brill waves}
\label{sec:moncrief:brill}

For the Brill waves of Section \ref{sec:indata:brill} we compute
\begin{subequations}
\begin{align}
   H^{\rm B}_{220} = & \sqrt{\frac{\pi }{5}}\frac{\Ab}{r^3} \Big[ \frac{e^{-(r/\sigma) ^2}}{3 \sigma ^2}   (4 r^5 + 2 r^3 \sigma ^2-3 r \sigma ^4)- \nonumber \\ 
   & ~~~~~~~~~~~ \sqrt{\pi } \frac{ \sigma ^3  }{2 } \erf\left(\frac{r}{\sigma }\right)\Big] \\
    h^{\rm B}_{120} = & 0 \\
    K^{\rm B}_{20} = & \sqrt{\frac{\pi}{5}}\frac{\Ab }{6  r^2}\Big[2  e^{-(r/\sigma) ^2} \left(2 r^2+3 \sigma ^2\right)- \nonumber \\ 
    & ~~~~~~~~~~ \frac{3 \sqrt{\pi}  \sigma ^3 }{r}\erf\left(\frac{r}{\sigma }\right)\Big] \\
    G^{\rm B}_{20} = & \sqrt{\frac{\pi }{5}} \frac{2 \Ab }{3 \sigma ^2}r^2 e^{- (r/\sigma)^2},
\end{align}
\end{subequations}
and then combine these functions to find  
\begin{subequations}
\begin{align}
    k^{\rm B}_{120} & =  -\sqrt{\frac{\pi }{5}} \frac{ \Ab }{6 r^3 \sigma ^4}  \left[ 3 \sqrt{\pi } \sigma ^7  \erf\left(\frac{r}{\sigma }\right) \right. +\\
    & ~~~~  \left.
    e^{-(r/\sigma) ^2} \left( 16 r^7 - 40 r^5 \sigma ^2 - 4 r^3 \sigma ^4  - 6 r \sigma ^6\right)\right] \nonumber \\
    k^{\rm B}_{220} & = \sqrt{\frac{\pi }{5}} \frac{ \Ab }{4 r^3 \sigma ^4} \left[ - 3 \sqrt{\pi } \sigma ^7  \erf\left(\frac{r}{\sigma }\right) \right. + \\
    & ~~~~  \left. e^{-(r/\sigma) ^2}\left(16 r^7 - 24 r^5 \sigma ^2 + 4 r^3 \sigma ^4 + 6 r \sigma ^6\right)\right]. \nonumber 
\end{align}
\end{subequations}

The gauge-invariant Moncrief function $R_{20}$ is then given by
\begin{align}
    R^{\rm B}_{20} & = \sqrt{\frac{\pi}{5}}\frac{\Ab}{12  r^2 \sigma ^2} \left[2  r e^{-(r/\sigma) ^2} \left(4 r^4+2 r^2 \sigma ^2+3 \sigma ^4\right)\right. -\nonumber \\
    & ~~~~~~~~~~ \left.3 \sqrt{\pi}  \sigma ^5 \erf\left(\frac{r}{\sigma }\right)\right].
\end{align}

\section{Expansion of $W_{\ell 0}$ in terms of spherical harmonics $Y_{\ell 0}$}
\label{app:wlm}

The functions $W_{\ell m}$ may also be written as
\begin{equation} \label{Wexp_Wlm}
    W_{\ell m} = \ell ( \ell + 1) Y_{\ell m} + 2 \partial^2_\theta Y_{\ell m}
\end{equation}
(see, e.g., eq.~D.12 in \cite{ShaB10}).  Since the second derivative of $Y_{\ell m}$ with respect to $\theta$ can be expressed in terms of spherical harmonics $Y_{\ell' m}$ with $\ell' = \ell - 2$, $\ell' = \ell - 4$ etc., we see that, for even (odd) $\ell$, the $W_{\ell m}$ can be written as a linear combination of all $Y_{\ell' m}$'s with even (odd) $\ell'$ satisfying $\ell \geq \ell' \geq m$.  In axisymmetry, i.e.~for $m = 0$, we can derive this linear combination from the properties of Legendre polynomials $P_\ell$, which are related to the axisymmetric spherical harmonics by
\begin{equation} \label{sph_harm}
    Y_{\ell 0} = \sqrt{\frac{2 \ell + 1}{4 \pi}} P_\ell.
\end{equation}

We start with the Legendre equation, which we may write in the form
\begin{equation}
    \frac{d^2 P_\ell}{d\theta^2} = - \frac{\cos \theta}{\sin \theta} \frac{dP_\ell}{d \theta} - \ell (\ell + 1) P_\ell 
    = x \frac{dP_\ell}{d x} - \ell (\ell + 1) P_\ell 
\end{equation}
where $x \equiv \cos \theta$ in the last step.  We then use the recurrence relation 
\begin{equation}
    x P'_\ell = P'_{\ell - 1} + \ell P_\ell
\end{equation}
(see, e.g., Eq.~12.25 in \cite{ArfW05}) to find
\begin{equation} \label{Ppp_expansion}
    \frac{d^2 P_\ell}{d\theta^2} 
    = P'_{\ell - 1}   - \ell^2 P_\ell 
\end{equation}
Now we can use the identity
\begin{equation} \label{eq:Leg_ident}
    P'_{n + 1} = P'_{n - 1} + (2n+1) P_n
\end{equation}
(see, e.g., 12.23 in \cite{ArfW05}) repeatedly. Starting with $n = \ell - 2$, eq.~(\ref{Ppp_expansion}) becomes
\begin{equation}
    \frac{d^2 P_\ell}{d\theta^2} 
    = P'_{\ell - 3} + (2 \ell - 3) P_{\ell - 2} - \ell^2 P_\ell, 
\end{equation}
next we use (\ref{eq:Leg_ident}) for $n = \ell - 4$ etc..  Starting with an even $\ell$, we at some point end up with a term $P'_3$, which we write as
\begin{equation}
    P'_3 = P'_1 + 5 P_2 = 5 P_2 + P_0,
\end{equation}
where we have used $P'_1 = 1 = P_0$.  We may therefore write \begin{equation}
        \frac{d^2 P_\ell}{d\theta^2} 
    = \sum _{n=0}^{\ell-2} (2n +1)P_n - \ell^2P_\ell~~~~~~~ (\mbox{$\ell > 2$ even, $n$ even}).
\end{equation}
 
Using (\ref{sph_harm}) again we then have
\begin{align}
    \partial^2_\theta\, Y_{\ell 0}
    & = -\ell^2Y_{\ell 0} + \sqrt{2\ell + 1} \sum_{n=0}^{\ell -2} \sqrt{2n +1} \, Y_{n0} \nonumber \\ 
    & ~~~~~~~~~~~~~~~~~~~ (\mbox{$\ell > 2$ even, $n$ even}),
\end{align}
 
which we may insert into (\ref{Wexp_Wlm}) to obtain 
\begin{align}
      W_{\ell 0}
    = \ell \, Y_{\ell 0} + 2\sqrt{2\ell + 1} \sum_{n=0}^{\ell -2} \sqrt{2n +1} \, Y_{n0}\nonumber \\ 
    ~~~~~~~(\mbox{$\ell > 2$ even, $n$ even}).
\end{align}
Since the function $q$ in (\ref{q_decomp}) contains only monopole and quadrupole terms, only the last two terms in this expansion,
\begin{equation} \label{eq:wl0_exp}
W_{\ell 0} = \ldots + 2 \sqrt{2 \ell + 1} \left( \sqrt{5} \, Y_{20} + Y_{00} \right)~~(\mbox{$\ell > 2$ even}),
\end{equation}
can yield a contribution in the integral (\ref{eq:Wlm_integral}) for $\ell > 2$.

\end{appendix}

\bibliography{references}

\end{document}